\documentclass[a4paper,twocolumn,11pt,accepted=2017-05-09]{quantumarticle}
\pdfoutput=1
\usepackage[utf8]{inputenc}
\usepackage[american]{babel}
\usepackage[T1]{fontenc}
\usepackage[pdftex]{graphicx}  
\usepackage[usenames,dvipsnames,svgnames]{xcolor}
 
\usepackage{dcolumn}
\usepackage{comment}
\usepackage{scalerel}
\usepackage{lipsum}
\usepackage{footnote}
\usepackage{amsmath}
\usepackage{amsfonts}
\usepackage{amssymb,bm}
\usepackage{color}
\usepackage[colorlinks,bookmarks=false,citecolor=NavyBlue,linkcolor=Red,urlcolor=blue]{hyperref}
\usepackage[export]{adjustbox}
\usepackage{amsthm}
\usepackage{simplewick}
\usepackage{bbold}
\usepackage{physics}
\usepackage{wasysym}
\usepackage[toc]{appendix}
\usepackage{mathrsfs}  
\usepackage{braket}
\usepackage{graphics}
\usepackage{todonotes}
\usepackage{multirow}
\usepackage{algorithm}
\usepackage{algpseudocode}
\usepackage{dsfont}
\usepackage{mdframed}

\newcommand{\Cliff}{\mathcal{C}} 

\newcommand{\Id}{\hat{\mathbb{1}}}
\newcommand{\PauliX}{{\hat{\text{X}}}}
\newcommand{\PauliY}{{\hat{\text{Y}}}}
\newcommand{\PauliZ}{{\hat{\text{Z}}}}

\newcommand{\MyTr}{\text{Tr}}
\newcommand{\parity}{{\hat{\mathcal{P}}}}
\newcommand{\Var}{\text{Var}}

\newcommand{\gaussgroup}{{\mathcal{U}_G}}
\newcommand{\gausstate}{{\mathcal{G}}}

\newcommand{\pf}{{\text{Pf}}}
\newcommand{\Haar}{{\text{Haar}}}
\newcommand{\nsamp}{{\mathcal{N}}}

\newcommand{\red}[1]{\textcolor{black}{ }\textcolor{black}{#1}}
\newcommand{\rednew}[1]{\textcolor{black}{ }\textcolor{black}{#1}}
\newcommand{\rednewnew}[1]{\textcolor{black}{ }\textcolor{black}{#1}}

\newcommand{\rednewthree}[1]{\textcolor{black}{ }\textcolor{black}{#1}}
\begin{document}
\title{The non-stabilizerness of fermionic Gaussian states}

\author{Mario Collura}
\affiliation{International School for Advanced Studies (SISSA), via Bonomea 265, 34136 Trieste, Italy}
\affiliation{INFN Sezione di Trieste, 34136 Trieste, Italy}
\author{Jacopo De Nardis}
\affiliation{Laboratoire de Physique Th\'eorique et Mod\'elisation, CNRS UMR 8089, CY Cergy Paris Universit\'e, 95302 Cergy-Pontoise Cedex, France}
\author{Vincenzo Alba}
\affiliation{Dipartimento di Fisica dell’Universit\`a di Pisa, 56127 Pisa, Italy}
\affiliation{INFN Sezione di Pisa, 56127 Pisa, Italy}
\author{Guglielmo Lami}
\affiliation{Laboratoire de Physique Th\'eorique et Mod\'elisation, CNRS UMR 8089, CY Cergy Paris Universit\'e, 95302 Cergy-Pontoise Cedex, France}

\begin{abstract}
We introduce an efficient method to quantify nonstabilizerness in fermionic Gaussian states, overcoming the long-standing challenge posed by their extensive entanglement. Using a perfect sampling scheme based on an underlying determinantal point process, we compute the Stabilizer Rényi Entropies (SREs) for systems with hundreds of qubits. Benchmarking on random Gaussian states with and without particle conservation, we reveal an extensive leading behavior equal to that of Haar random states, with logarithmic subleading corrections. We support these findings with analytical calculations for a set of related quantities, the participation entropies in the computational (or Fock) basis, for which we derive an exact formula. We also investigate the time evolution of \red{non-stabilizerness} in a random unitary circuit with Gaussian gates, observing that it converges in a time that scales logarithmically with the system size. Applying the sampling algorithm to a two-dimensional free-fermionic topological model, we uncover a sharp transition in \red{non-stabilizerness} at the phase boundaries, highlighting the power of our approach in exploring different phases of quantum many-body systems, even in higher dimensions. 
\end{abstract}

\maketitle

\paragraph{Introduction. ---} 
Quantum states of free fermionic particles play a crucial role in condensed matter physics~\cite{Fradkin_2013} and quantum chemistry~\cite{Szabo_1996,Mcardle_2020}, serving as foundational tools for understanding the quantum phases of matter and enabling key computational techniques~\cite{Giuliani_Vignale_2005}. These fermionic `Gaussian' states are fully characterized by their two-points correlation functions, and combine the ability to capture essential physical features with analytical tractability. In the last few decades, they have been used as benchmarks for inspecting paradigmatic properties of many-body systems, ranging from quantum thermalization~\cite{Rigol_2007,Essler_2016,Lucas_2023,Bejan_2024_2} to disorder induced localization~\cite{PhysRev.109.1492}, disordered systems~\cite{PhysRevLett.70.3339}, topological features~\cite{RevModPhys.83.1057}, entanglement spreading~\cite{Calabrese_2005, essler2016quench,Fagotti_2008,Alba_2017,alba2021generalized,alba2021spreading,caceffo2024fate}.
In addition, Gaussian states have been considered in the realm of quantum computation, where they
constitute an important class of computational states~\cite{Weedbrook2012,Yurke1986,Lami_2018}. Free-fermionic circuits, studied under the names `matchgate' circuits~\cite{Valiant_2001, Terhal_2002, Jozsa_2008, Mocherla_2024, Dias_2024} and fermionic linear optics~\cite{Bravyi_2004, Reardonsmith_2024}, have been found to be classically simulable, yet capable of exhibiting non-trivial features~\cite{Bittel_2024,Mele_2024,Oszmaniec_2022,Adhikari_2024,Hackl_2018,Vanhala_2024}. \\
In these works, quantum entanglement has played a central role, being recognized as a fundamental feature of quantum states that sets them apart from the classical world. However, entanglement alone does not guarantee, for instance, classical un-simulability, as fermionic Gaussian states can possess arbitrary entanglement while remaining efficiently classically simulable. In general, quantum states can possess a range of distinctive characteristics that contribute to their intrinsic complexity, extending beyond just entanglement. The rigorous mathematical framework for defining these `quantum resources' is a well-established part of quantum information theory~\cite{Gour_2024, Chitambar_2019}. \\
In particular, the quantity known as \textit{nonstabilizerness}, or quantum magic, has emerged as a fundamental resource of the quantum world~\cite{Howard_2017, Winter_2022}. At its heart, the idea is that the states generated by unitary transformations in the Clifford group, known as stabilizer states, are not only simulable using efficient classical algorithms~\cite{Gottesman_1997, Gottesman_1998_1, Gottesman_1998_2} but also easy to produce in the laboratory. This arises because Clifford operations can be implemented fault-tolerantly, as they are compatible with quantum error correction codes~\cite{Gottesman_1997,Zeng_2011,Veitch_2014}. On the other hand, non-Clifford gates are essential for enabling universal quantum computation~\cite{Bravyi_2005} but they are challenging to implement and thus can be regarded as a sort of expensive resource~\cite{Winter_2022}. \red{Non-stabilizerness}, therefore, quantifies the intrinsic amount of non-Clifford resources required to prepare a state, and is also closely related to the costs associated with classical simulation algorithms. \\
Recently, \red{non-stabilizerness} has become a focus of study in many-body physics, offering valuable insights into the properties of different phases of matter in spin systems~\cite{Oliviero_2022,Lami_2023_2,Tarabunga_2024,Rattacaso_2023,Passarelli_2024_1,Passarelli_2024_2}, frustrated and topological models~\cite{Kitaev_2006,Odavic_2023,Catalano_2024,Frau_2024_1,Delafuente_2024}, lattice gauge theories~\cite{Tarabunga_2023_1,Falcao_2024}, conformal theories~\cite{Frau_2024_2,Fan_2024}, 
measurements induced criticality~\cite{Paviglianiti_2024,Bejan_2024_1,Fux_2024},
and even high-energy physics~\cite{Robin_2024,Brokemeier_2024,Chernyshev_2024}.\\
Despite these advances, the nonstabilizerness properties of fermionic Gaussian states remain unexplored, likely due to the absence of suitable analytical or computational tools to probe this feature. In fact, many measures of magic require expensive minimization procedures~\cite{Winter_2022}, limiting their applicability to systems with only a small number of qubits~\cite{Sarkar_2020}. While the Stabilizer Rényi Entropies (SREs)~\cite{Leone_2022,Leone_2024} are instead defined through simple operator expectation values, they also incur a computational cost that scales exponentially with system size, generally making them impractical for systems larger than a few dozen qubits.
Only when the state under investigation has limited entanglement, allowing for a faithful description in terms of a Matrix Product State (MPS), tailored techniques may be employed to simplify the computation of SREs~\cite{Haug_2023_1,Lami_2023_2,Lami_2024,Tarabunga_2024_2,Tarabunga_2024}. However, fermionic Gaussian states typically exhibit extensive entanglement~\cite{Bianchi_2021}, placing them in a class of states for which no efficient method has been available to quantify nonstabilizerness.

In this work, we address this challenge by presenting the \textit{first detailed investigation of nonstabilizerness in fermionic Gaussian states}. First, by leveraging a novel perfect sampling algorithm for the underlying Determinantal Point Process~\cite{Scardicchio_2009, Kulesza_2012}, we address the computation of SREs. This new method allows us to evaluate \red{non-stabilizerness} in free-fermionic systems containing hundreds of qubits. Afterwards, we employ our approach by studying the behavior of SREs in random Gaussian states, both with and without $U(1)$ symmetry. We compare our findings with analytical results for a similar quantity, namely the Participation Rényi Entropies (PREs) over the computational basis, for which we find an analytical formula.
We also study the time evolution of magic in a random unitary circuit with local Gaussian gates. Finally, we apply our technique to compute the non-stabilizerness of the ground state of a two-dimensional non-interacting fermionic topological superconductor, observing a notable shift in \red{non-stabilizerness} properties near the critical point of the phase transition. This suggests that \red{non-stabilizerness} is closely tied to the critical behavior of the system, with potential implications for the properties of the resulting (topological) phase.
\red{Finally, we emphasize that although other resource theories—such as the resource theory of non-Gaussianity~\cite{Hebenstreit_2019, Dias_2024, Najafi_2024, Marian_2013, Lumia_2024}—may be relevant for fermionic models, our focus here is exclusively on their non-stabilizerness. The applicability and usefulness of a particular resource theory for a certain many-body problem depends crucially on the quantum platform used for its implementation. Specifically, when a fermionic system is realized natively, non-Gaussianity is the key resource; in contrast, when the system is mapped onto qubits and implemented on a digital quantum platform—which is arguably the most relevant scenario for currently available quantum technologies advancing— the resource theory of non-stabilizerness provides the appropriate framework.}

\paragraph{Preliminaries. ---} 
We consider a quantum system of $L$ qubits, with total Hilbert space dimension $D=2^L$. The local Pauli operators are denoted as $\hat{P}_i \in \{\Id, \PauliX, \PauliY, \PauliZ\}$ for $i=1,2... L$, while $\hat{\pmb{P}} \in \{\Id, \PauliX, \PauliY, \PauliZ \}^{\otimes L}$ represents a generic tensor product of Pauli operators (i.e.\ a Pauli string). The set $\{ \frac{\hat{\pmb{P}}}{\sqrt{D}} \}$ forms an orthonormal basis for the space of operators, since $\MyTr[\hat{\pmb{P}}' \hat{\pmb{P}}] = D \delta_{\pmb{P}' \pmb{P} }$. Majorana operators are defined using the standard Jordan-Wigner mapping, as follows:
\begin{eqnarray*}
\hat{\gamma}_{2i-1} & = & \PauliZ_1 ... \PauliZ_{i-1} \PauliX_{i} \Id_{i+1} ... \Id_{L}\, , \\ 
\hat{\gamma}_{2i} & = & \PauliZ_1 ... \PauliZ_{i-1} \PauliY_{i} \Id_{i+1} ... \Id_{L} \, ,
\end{eqnarray*}

where $i=1,2... L$. The $2L$ Majorana operators are Hermitian and satisfy the canonical anti-commutation relations $\{ \hat{\gamma}_{\mu}, \hat{\gamma}_{\nu} \} = 2 \delta_{\mu \nu}$, for $\mu, \nu \in \{1,2, ... \, 2L \}$. They can also be written in terms of the fermionic creation annihilation operators $\hat{c}_i, \hat{c}_i^{\dag}$, as $\hat{\gamma}_{2i-1} = (\hat{c}_i^{\dag} + \hat{c}_i)$, $\hat{\gamma}_{2i} = i(\hat{c}_i^{\dag} -\hat{c}_i)$. The computational basis is identified with the set of Fock states $\ket{\pmb{z}} = (\hat{c}^{\dag}_1)^{z_1} ... (\hat{c}^{\dag}_L)^{z_L} \ket{0 ...0}$, where $z_i \in \{ 0, 1\}$ is the local occupation-number and $\ket{0...0}$ is the the Fermi vacuum. Majorana monomials are generic strings of Majorana operators $\hat{\gamma}^{\pmb{x}} = (\hat{\gamma}_{1})^{x_1} (\hat{\gamma}_{2})^{x_2} ... (\hat{\gamma}_{2L})^{x_{2L}}$, where $\pmb{x} \in \{0,1\}^{2L}$ is a binary vector whose components indicate which Majorana operators are present (or not) in the monomial. Majorana monomials are in one to one correspondence with Pauli strings, forming a complete orthogonal basis as well, since $\MyTr[(\hat{\gamma}^{\pmb{x}'})^{\dag} \hat{\gamma}^{\pmb{x}}] = D \delta_{\pmb{x}' \pmb{x}}$. A unitary operator $\hat{U}$ is Gaussian if and only if it acts on the Majorana operators by rotating the vector of $\hat{\gamma}_{\mu}$ as
\begin{equation}\label{eq:majorana_rotation}
    \hat{U}^{\dag} \hat{\gamma}_{\mu} \hat{U} = \sum_{\nu} O_{\mu \nu} \hat{\gamma}_{\nu} \, ,
\end{equation}
with $O \in SO(2L)$ a real orthogonal matrix, corresponding to a Bogoliubov transformation~\cite{Bloch_1962}. The Gaussian unitary group, denoted as $\gaussgroup$, is generated by the algebra of Majorana fermion pairs $\hat{\gamma}_{\mu} \hat{\gamma}_{\nu}$. Eq.~\eqref{eq:majorana_rotation} implies that all Gaussian operators commute with the parity operator $\parity = (-i)^L \hat{\gamma}_1 \hat{\gamma}_2 ... \hat{\gamma}_{2 L} = \PauliZ_{1}  ... \PauliZ_{L}$.
The set of Gaussian states is defined as~\cite{Bravyi_2004}: $\gausstate_L = \{ \hat{\rho} = \hat{U}^{\dag} \bigotimes_{i=1}^L \left( \frac{\Id + \lambda_i \PauliZ}{2} \right) \hat{U} \, , \, \, \hat{U} \in \gaussgroup  \ , \lambda_i \in [-1, +1] \}$. It includes also non pure states, while setting all the coefficients $\lambda_i$ to $1$ restricts the ensemble to pure Gaussian states $\gausstate_L^{\text{pure}} = \{ \hat{U} \ket{0...0} \, , \hat{U} \in \gaussgroup \}$, which are generated from the vacuum by the action of a Gaussian operator. Gaussian states are fully characterized by the real and skew-symmetric covariance matrix~\cite{Surace_2022}
\begin{equation}
\Gamma_{\mu \nu}(\hat{\rho}) = -\frac{i}{2} \MyTr([\hat{\gamma}_{\mu}, \hat{\gamma}_{\nu}] \hat{\rho}) \, ,
\end{equation} 
which contains all two-points Majorana correlation functions. When applying $\hat{U} \in \gaussgroup$ to a state $\hat{\rho}$, the covariance matrix rotates accordingly as $\Gamma(\hat{U}^{\dag} \hat{\rho} \hat{U}) = O \Gamma O^T$. $\Gamma$ can be readily expressed in terms of the standard fermionic correlation matrix $C_{\mu \nu} (\hat{\rho})=\MyTr(\hat{\mathbb{c}}_{\mu}^{\dag} \hat{\mathbb{c}}_{\nu} \hat{\rho})$, where $\hat{\mathbb{c}}$ is defined by
$\hat{\mathbb{c}}_{2i-1}=\hat{c}_i$, $\hat{\mathbb{c}}_{2i}=\hat{c}_i^{\dag}$. In particular, $\Gamma_{\mu \nu}(\hat{\rho}) = -i \left( 2 \Omega^*  C(\hat{\rho}) \Omega^T - \mathbb{1} \right) _{\mu \nu}$ where $\Omega = \bigoplus_{i=1}^{L} \Omega^{(i)}$ is the unitary matrix that transforms the $\hat{\mathbb{c}}$ into $\hat{\gamma}$~\cite{Surace_2022}. For the vacuum state $\ket{0...0}$, the covariance matrix reads
\begin{equation*}
\Gamma_0 = \Gamma(\ket{0...0}) = \bigoplus_{i=1}^L \begin{pmatrix}
    0 & 1 \\
    -1 & 0 \\
\end{pmatrix} \, .
\end{equation*}
The celebrated Wick's theorem enables the calculation of any correlation function in terms of two-point correlation functions, and can be formulated as follows~\cite{Surace_2022}
\begin{equation}
    \MyTr(\hat{\rho} \, \hat{\gamma}^{\pmb{x}}) = i^{\frac{|\pmb{x}|}{2}} \pf[\Gamma|_{\pmb{x}}] \, ,
\end{equation}
where $\Gamma|_{\pmb{x}}$ denotes the square sub-matrix of $\Gamma$ that includes all rows and columns associated with indices equal to 1 in $\pmb{x}$, and $|\pmb{x}|$ represents the total count of such indices. Finally, we define particle-number-preserving Gaussian transformations as those that commute with the total particle number operator $\hat{N} = \sum_{i=1}^L \hat{c}^{\dag}_i \hat{c}_i$. Observe that any particle number-preserving transformation must map the vacuum state $\ket{0...0}$ into itself. This imposes the condition $O \Gamma_0 O^T = \Gamma_0$, where $O$ is the orthogonal transformation corresponding to $U$. In fact, it can be easily show that $O \in O(2L)$ represents a number preserving transformation iff it commutes with $\Gamma_0$, meaning $O$ is symplectic~\cite{Mele_2024}. The set of pure fermionic Gaussian states with $N$ particles is $\gausstate_{L,N}^{\text{pure}} = \{ \hat{U} \ket{\pmb{z}} \, , \hat{U} \in \gaussgroup, \, [\hat{U},\hat{N}]=0, \, \sum_{i=1}^L z_i = N \}$.\\

\paragraph{Characteristic distribution and Stabilizer Rényi Entropies (SREs). ---}
Given an Hermitian operator $\hat{O}$ and a Pauli string $\hat{\pmb{P}}$, we define
\begin{equation}\label{eq:prob}
   \pi_{O}(\pmb{P}) \equiv \frac{1}{D} \frac{\MyTr [\hat{\pmb{P}} \hat{O}]^2}{\MyTr[\hat{O}^2]} \, .
\end{equation}
Over all Pauli strings, $\pi_{O}$ forms a length-$D^2$ vector with elements summing to 1. In fact, completeness of Pauli operators gives $D^{-1} \sum_{\pmb{P}} (\MyTr[\hat{\pmb{P}} O])^2 = \MyTr[\hat{O}^2]$, where $\MyTr[\hat{O}^2]$ normalizes Eq.~\eqref{eq:prob}. Thus, $\pi_{O}$ is a probability distribution, often called characteristic function~\cite{Gross_2016,Leone_2022}, capturing the overlap of $O$ with all Pauli operators~\cite{Turkeshi_2023_2}.
The evolution of this distribution under generic many-body time dynamics has been investigated in the context of operator spreading, where an initially localized operator $\hat{O}_0$ gains overlap with Pauli strings of increasing length~\cite{Khemani_2018, Nahum_2018, Qi_2019}. Marginalized versions of $\Pi$ are often used, such as distributions of operator lengths or the fraction of strings ending at position $i$. \\

\noindent
We now examine the characteristic distribution of a fixed state $\hat{\rho}$ (which might be mixed). For $\alpha > 0$, the $\alpha$-Stabilizer Rényi Entropy (SRE) is defined as~\cite{Leone_2022}
\begin{equation}\label{eq:sre}
    M_{\alpha}(\rho) = \frac{1}{1-\alpha} \log \sum_{\pmb{P}} \pi_{\rho}^{\alpha}(\pmb{P}) - \log D \, , 
\end{equation}
and, excluding an additive constant, coincides with the $\alpha$-Rényi entropy of the distribution $\pi_{\rho}$. For $\alpha \to 1$, the SRE reduces to the Shannon entropy: $M_{1}(\rho) = - \sum_{\pmb{P}} \pi_{\rho}(\pmb{P}) \log \pi_{\rho}(\pmb{P}) - \log D$. For all $\alpha$, 
the SREs are trivially upper bounded by $\log D$, which however is in general a loose bound for pure states~\cite{Leone_2022}. In addition to being a natural way of characterizing the distribution $\pi_{\rho}$~\eqref{eq:prob}, SREs have also proven to have substantial relevance in the context of quantum information as measures of nonstabilizerness, aka magic. In a nutshell, nonstabilizerness is the resource required for quantum states to be unattainable through simple Clifford circuits (plus Pauli measurements). The latter are regarded as easy to implement ``free'' operations, while non-Clifford gates are resources. Recall that the Clifford group $\Cliff_L$ includes the unitary transformations $\hat{U}_c$ that map Pauli strings into Pauli strings under conjugation, i.e. $\hat{U}_c^{\dag} \hat{\pmb{P}} \hat{U}_c = e^{i \theta \frac{\pi}{2}} \hat{\pmb{P}}'$ ($\theta \in \{0,1,2,3\}$) for all $\hat{\pmb{P}}$, where $e^{i \theta \frac{\pi}{2}}$ is an irrelevant phase. Pure stabilizer states, denoted as STAB, are those derived from the reference state 
$\ket{0 ... 0}$ through Clifford transformations, and therefore they carry no resource. SREs are widely regarded as effective measures of nonstabilizerness, particularly for $\alpha \geq 2$, where their monotonicity is rigorously established~\cite{Leone_2024,Haug_2023_2}. Besides, SREs provide useful bounds for other measures of magic~\cite{Leone_2024}. \\

In Ref.~\cite{Turkeshi_2023_2}, it a slightly modified version of the SREs, named \textit{Filtered Stabilizer Rényi Entropies} (filtered SREs), has been introduced in which the contribution of the identity $\hat{I}=\Id^{\otimes L}$ is removed from the sum in Eq.~\eqref{eq:sre}. This prevents dominance of this contribution for large $L$ and $\alpha > 2$, enabling filtered SREs to distinguish typical from atypical states, unlike standard SREs. \\

A powerful method for evaluating SREs involves sampling in the Pauli basis~\cite{Lami_2023_2,Lami_2024}. In fact, the argument of the logarithm in Eq.~\eqref{eq:sre} can be rewritten as $\sum_{P} \pi_{\rho}(\pmb{P}) \pi_{\rho}^{\alpha-1}(\pmb{P})$, which represents the expected value of $\pi_{\rho}^{\alpha-1}$. Consequently, the SREs can be estimated by sampling strings $\pmb{P}$ with probability $\pi_{\rho}(\pmb{P})$, and averaging $\pi_{\rho}^{\alpha-1}(\pmb{P})$ \red{($\log \pi_{\rho}(\pmb{P})$ for the case $\alpha=1$)} over these samples (\red{a detailed description of the averaging procedure is provided in Appendix}~\ref{appendix1}). 
\rednew{The statistical estimator obtained in this way clearly provides an unbiased estimate of the argument inside the logarithm in Eq.~\eqref{eq:sre} for $\alpha > 1$, from which one can compute $M_{\alpha}$. For the special case $\alpha = 1$, the estimator directly yields $M_1$. It is crucial to characterize how the fluctuations (variance) of the estimator scale with the relevant parameters. First, since the estimation is obtained through a standard sample average, the fluctuations decrease proportionally to the inverse square root of the number of samples.} \red{Hoverer, as discussed in Ref.~\cite{Lami_2023_2}, the number of samples required to estimate $M_{\alpha}$ with a given accuracy depends on the Renyi index $\alpha$. Specifically, for $\alpha = 1$, the number of required samples scales at most as $L^2$, demonstrating that the Shannon stabilizer entropy can be efficiently estimated through sampling. In contrast, for $\alpha > 1$, the number of required samples scales exponentially with the system size $L$ in the worst-case scenario. Although this exponential scaling could, in principle, pose significant limitations, in practice, growth has been observed to be relatively mild~\cite{Lami_2023_2,Lami_2024}. Moreover, since sampling is a highly parallelizable task, it is possible to significantly increase the number of samples using parallel computation. We refer to Ref.~\cite{Lami_2023_2} for further discussions on the efficiency of the sampling.} \\
While \red{the sampling approach is general}, it has been successfully applied to cases where the state $\ket{\psi}$ is an MPS, for which an exact and efficient sampling is possible~\cite{Lami_2023_2,Lami_2024}. In this case, alternative methods for computing SREs, primarily based on the replica approach~\cite{Haug_2023_1,Haug_2023_2,Tarabunga_2024}, are also available. However, MPS have fundamental limitations, specifically in the amount of entanglement they can capture, which makes it challenging to apply these methods to study the many-body dynamics or 2D systems. Therefore, developing methods to compute the magic of a state, independent of its entanglement, is of paramount importance. In the following, we tackle this challenge by applying the sampling estimation to fermionic Gaussian states, for which we develop a novel, ad hoc sampling algorithm. \\
\rednewthree{Before of detailing this sampling scheme, let us however describe a fundamental limitation of the method. Recently, it has been observed that there exists a class of states, known as \emph{pseudo-magic states}, which possess only a limited amount of magic (for instance, scaling logarithmically with the system size) yet are indistinguishable from highly magic states (e.g.\ with magic scaling linearly in the system size) for any computationally bounded observer~\cite{Gu_2024}. In particular, no measurement scheme implementable in polynomial time can distinguish between these two ensembles. Since the proposed sampling from $\pi_{\rho}(\pmb{P})$ is such a polynomial-time procedure, it follows that this protocol cannot discriminate efficiently pseudo-magic states from genuinely magic ones. Nevertheless, the physical relevance of the notion of pseudo-magic, or more generally
of pseudo-resources~\cite{aaronson2023quantumpseudoentanglement}, remains an open
question. Indeed, many of the currently known constructions (e.g.\ subset-phase-states) are unstructured and do not incorporate fundamental physical
principles such as locality, finite energy, etc.}

\paragraph{SREs for Fermionic Gaussian states. ---}
We consider the problem of accessing and characterizing the probability distribution $\pi_{\rho}(\pmb{P})$ of a fermionic Gaussian state $\rho$, given its known covariance matrix $\Gamma$. For instance, one might be interested in computing the SREs $M_{\alpha}(\rho)$ associated with the distribution. More generally, one might seek to explore the distribution $\pi_{\rho}(\pmb{P})$ itself. First, we notice that since there exist a one-to-one mapping between Majorana monomials and Pauli strings, the distribution $\pi_{\rho}(\pmb{x}) = D^{-1} |\MyTr[\hat{\rho} \gamma^{\pmb{x}}]|^2 / \MyTr[\hat{\rho}^2]$ is perfectly equivalent to $\pi_{\rho}(\pmb{P})$. 
Now, using the Wick theorem and the fact that the square of the Pfaffian equals the determinant, we obtain 
\begin{equation}\label{eq:dpp_1}
   \pi_{\rho}(\pmb{x}) = \frac{\det[\Gamma|_{\pmb{x}}]}{\det[\mathbb{1} + \Gamma]} \, ,
\end{equation}
where the denominator is a known expression for the purity of a fermionic Gaussian state~\cite{Surace_2022}. Eq.~\eqref{eq:dpp_1} shows that the distribution $\pi_{\rho}$ defines a Determinantal Point Processes (DPP) over a system of $2L$ bits, whose configurations are labeled by $\pmb{x}$. DPP are a prominent class of probabilistic models, in which the probability of a given configuration is represented by the determinant of a sub-matrix —specifically, a principal minor—of a fixed kernel matrix $K$~\cite{Scardicchio_2009, Kulesza_2012,Najafi_2024}. DPP were initially developed to describe the distribution of fermions in thermal equilibrium~\cite{Macchi_1975}, and later applied across a remarkable range of fields. In particular, they find significant applications in random matrix theory and machine learning~\cite{Kulesza_2012}. The most extensively studied case is when $K$ is positive semi-definite, a setting for which well-established sampling strategies are available~\cite{Hough_2006,Scardicchio_2009}. However, more general scenarios in which $K$ is non-symmetric can also be considered, since any matrix whose principal minors are nonnegative defines a DPP~\cite{Gartrell_2020}. In Eq.~\eqref{eq:dpp_1}, for instance, the kernel is the real skew-symmetric matrix $\Gamma$, implying positive correlations, while standard DPPs are limited to negative correlations~\cite{Gartrell_2020}.
In the following Section, we consider the DPP Eq.~\eqref{eq:dpp_1} and we introduce a novel, efficient method to sample configurations $\pmb{x}$, i.e.\ Majorana monomials, accordingly. Our method differs from standard Monte Carlo approaches, as it does not rely on a Markov chain. It is instead a perfect sampler, meaning that our Algorithm can sample directly from Eq.~\eqref{eq:dpp_1}, with zero autocorrelation time. Technically, our method closely resembles the iterative approach introduced in Ref.~\cite{Launay_2020} for generic (non-symmetric) DPPs. \\
By using this sampling algorithm, one can analyze the distribution $\pi_{\rho}$ and estimate the SREs as described above. For practical reasons, we adopt a filtered version of the SREs, excluding the contributions of the identity $\hat{I}=\Id^{\otimes L}$ and parity operators $\parity = (-i)^L \gamma_1 \gamma_2 ... \gamma_{2 L} = \PauliZ_{1}  ... \PauliZ_{L}$, which are trivial for pure fermionic Gaussian states ($\braket{\psi|\hat I|\psi}^2=\braket{\psi|\parity|\psi}^2=1$, for all $\ket{\psi} \in \gausstate_L^{\text{pure}}$). In particular, we define 
\begin{equation}\label{eq:fsre}
    \tilde{M}_{\alpha}(\rho) = \frac{1}{1-\alpha} \log \sum_{\pmb{P}} \tilde{\pi}_{\rho}^{\alpha}(\pmb{P}) - \log (D -2) \, , 
\end{equation}
with the probability distribution $\tilde{\pi}$ defined by
\begin{equation}\label{eq:fsre_prob}
    \tilde{\pi}_{\rho}(\pmb{P}) \propto \begin{cases}
        0 \text{   if } \hat{\pmb{P}}= \hat{I}, \parity \\
        \pi_{\rho}(\pmb{P}) \text{   otherwise} \\
    \end{cases}
\end{equation}
The filtered SREs in Eqs.~\eqref{eq:fsre} and~\eqref{eq:fsre_prob} are motivated, as in Ref.~\cite{Turkeshi_2023_2}, by the need to remove trivial contributions from the operatorial entropies, which lead to anomalously large weights for typical states. The only difference with Ref.~\cite{Turkeshi_2023_2} is that, for pure fermionic Gaussian states, we also remove the parity operator, which is a trivial symmetry of these states.\rednewnew{The filtered SREs in Eq.\ref{eq:fsre} inherit the monotonicity properties of conventional SREs when restricted to parity-preserving operations (even though, in principle, they could allow for a broader set of free operations than the standard SRE).
Details on the sample estimation of the filtered SREs and the associated sources of errors are provided in the Appendix~\ref{appendix1}.} \\

\paragraph{Majorana Sampling. ---} 
Sampling from the set of Pauli strings $\hat{\pmb{P}}$, or equivalently from the set of configurations $\pmb{x}$, with a size of $D^2=2^{2L}$, may seem exponentially difficult at first glance. To overcome this challenge, we express the full probability in terms of conditional and prior (or marginal) probabilities as follows:
\begin{equation}\label{eq:chain_prob}
\pi_{\rho}(\pmb{x})
= \pi_{\rho}(x_1)
\pi_{\rho}(x_2|x_1)
\cdots
\pi_{\rho}(x_{2L}|x_1\cdots x_{2L-1})  \, ,
\end{equation}
where $\pi_{\rho}(x_{\mu}|x_{1}\cdots x_{\mu-1})$ denotes the probability that bit $\mu$ assumes the value $x_{\mu}$, given that the string $x_1 \cdots x_{\mu-1}$ has already occurred at positions $1 \dots \mu-1$, no matter the occurrences in the rest of the system (i.e.\ marginalising over all possible strings for the bits $\mu+1 \dots 2L$). Thanks to these observations, one can generate outcomes and their associated probabilities by sequentially iterating over each binary variable and sampling each based on its respective conditional probability. The crucial point now is  to find an efficient way to calculate the conditional probability at a generic step $\mu$ of the iterative process, which reads $\pi_{\rho}(x_{\mu}|x_{1}\cdots x_{\mu-1}) = \frac{\pi_{\rho}(x_1\cdots x_{\mu})}{\pi_{\rho}(x_1\cdots x_{\mu-1})}$. We notice that (see~\eqref{eq:dpp_1})
\begin{align}
    \begin{split}
        &\pi_{\rho}(x_1\cdots x_{\mu}) = \sum_{x_{\mu+1}' ... x_{2L}'} \pi_{\rho}(x_{1} ... x_{\mu} , x_{\mu+1}' ... x_{2L}')= \\ &= (\det[\mathbb{1} + \Gamma])^{-1} \sum_{\pmb{x}'} \delta_{x_1 x_1'} ... \delta_{x_{\mu} x_{\mu}'} \,
  \det[\Gamma|_{\pmb{x}'}] \, , 
    \end{split}
\end{align}
which essentially sums the determinants of all sub-matrices of $\Gamma$, with the constraint to include all rows (columns) corresponding to indices $x_1, \ldots, x_{\mu}$ that have been fixed to 1. Interestingly, a known formula exists for this quantity~\cite{Launay_2020,Kulesza_2012} and is given by 
\begin{align}\label{eq:formdet}
    \begin{split}
        \pi_{\rho}(x_1\cdots x_{\mu}) = \frac{\det[(\pmb{1}_{[\mu+1,2L]} \,  + \, \Gamma)|_{(x_1\cdots x_{\mu},1...1)}]}{\det[\mathbb{1} + \Gamma]} \,, 
    \end{split}
\end{align}
where $\pmb{1}_{[\mu+1,2L]}$ is the diagonal matrix with $1$ in the interval $[\mu+1,2L]$ and $0$ otherwise. Notably, in the case $\mu=0$, the numerator simplifies to $\det[\mathbb{1} + \Gamma]$, giving the correct normalization of $\pi_{\rho}$. Eq.~\eqref{eq:formdet} provides an efficient way to calculate partial probabilities, and together with Eq.~\eqref{eq:chain_prob}, defines a simple iterative algorithm that we summarize in the table~\ref{alg:QA}.

\begin{algorithm}[H]
\caption{Majorana sampling of Gaussian States}\label{alg:QA}
\begin{flushleft}
\hspace*{\algorithmicindent} \textbf{Input}: the $2L \times 2L$ covariance matrix $\Gamma$ of the state
\end{flushleft}
\begin{algorithmic}[1]
\State Compute the normalization factor (purity): $\det[\mathbb{1} + \Gamma]$ 
\State Initialize $\pmb{x}=()$, and $\Pi=1$
\For{($\mu=1$, $\mu=2L$, $\mu++$)}
     \State Compute 
     $\pi_{\rho}(\pmb{x}, x_{\mu})$, as in Eq.~\eqref{eq:formdet}, and \indent
      $\pi_{\rho}(x_{\mu}|\pmb{x}) = \Pi^{-1} \pi_{\rho}(\pmb{x}, x_{\mu})$ for $x_{\mu}\in \{ 0,1 \}$
     \State  Set $x_{\mu}$ to $0$ or $1$ randomly with probability \indent $\pi_{\rho}(x_{\mu}|\pmb{x})$
     \State Update $\Pi \rightarrow \Pi \cdot \pi_{\rho}(x_{\mu}|\pmb{x})$ and $\pmb{x} \rightarrow (\pmb{x}, x_{\mu})$
\EndFor
\end{algorithmic}
\begin{flushleft}
\hspace*{\algorithmicindent} \textbf{Output}: a string $\pmb{x} \in \{0,1\}^{2L}$, the \indent probability $\pi_{\rho}(\pmb{x})$
\end{flushleft}
\end{algorithm}

Note that the core of the algorithm involves evaluating $2L$ determinants of sub-matrices whose sizes range from $2L$ to $1$. This task can be performed at a computational cost of $o(L^4)$, which thus represents the overall cost per sample. In practice, determinants can be evaluated efficiently using optimized linear algebra routines, and samples can be extracted in parallel, making the algorithm highly efficient. \\

\begin{figure}[t!]
\centering
\includegraphics[width=0.9\linewidth]{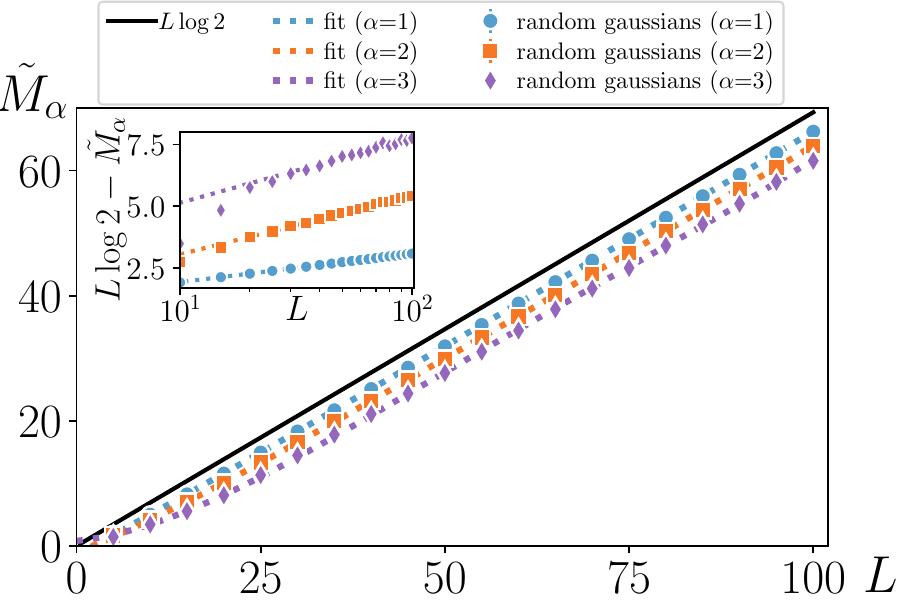}
\caption{Average filtered SREs $\tilde{M}_{\alpha}$ ($\alpha=1,2,3$) for random free-fermionic states, as a function of the system size $L$ \red{(particle-number is not conserved)}. The black line represents the leading extensive term for generic (non-Gaussian) Haar states, which is $L \log 2$. 
\red{We averaged over $200$ realizations of random states in $\gausstate_{L}^{\text{pure}}$, with $5 \cdot 10^3$ Majorana samples for each state}. Inset: the difference $L \log 2 - M_{\alpha}$.
\label{fig:random_gaussian_1} }
\end{figure}

\paragraph{Random Gaussian states. ---}
We first consider the case of random pure fermionic 
Gaussian states \red{without particle-number conservation}, which are defined by the covariance matrix $\Gamma = O \Gamma_0 O^T$, with the matrix $O$ taken from the Haar (i.e.\ uniform) distribution over the orthogonal group $O(2L)$. These states, regarded as typical eigenstates of free Hamiltonians~\cite{Lydzba_2020, Lydzba_2021}, have garnered considerable interest, with numerous studies aimed at characterizing their (average) entanglement~\cite{Bianchi_2021}.
Here we instead study the nonstabilizerness of such states. To this purpose, we generate a set of random pure Gaussian states and use the Algorithm~\ref{alg:QA} to estimate their filtered SREs $\tilde{M}_{\alpha}$, with $\alpha = 1,2,3$. Results are shown in Figure~\ref{fig:random_gaussian_1} for system of size $L \in [5,100]$. Notice, that for random Gaussian states the entanglement entropy $S_E(l) = - \MyTr[ \rho_l \log \rho_l]$, where $\rho_l$ denotes the reduced density matrix of the first $l$ sites, is extensive~\cite{Bianchi_2021}, meaning that simulating with MPS techniques typical fermionic states of these sizes is prohibitive. For instance, at $L=100$, the observed average half-chain entanglement entropy $S_E(L/2) \simeq 19.4$ requires an MPS bond dimension of $\chi \sim o(\exp(S_E(L/2))) \sim o(10^8)$. In contrast, our technique, which operates directly within the covariance matrix formalism, is entirely unaffected by the state's entanglement. For each value of $L$, we utilized $400$ realizations of random Gaussian states and employed $5 \cdot 10^3$ Majorana samples per realization to estimate the filtered SREs. 

Qualitatively, the average values of $\tilde{M}_{\alpha}$ are close to the entropies upper bound $\log D = L \log 2$ (black line). This also corresponds to the leading extensive term in the filtered SREs for generic (non-Gaussian) Haar-random states~\cite{Turkeshi_2023_2}. The deviation $L \log 2 - \tilde{M}_{\alpha}$ from this value (see inset) appears to grow as $\sim \log L$, thus giving only a sub-leading correction to the Haar scaling. 
In particular, fitting the relation $L \log 2 - \tilde{M}_{1} \simeq a_{\alpha} \cdot \log L + b_{\alpha}$ returns the coefficients: \rednewnew{$a_{\alpha} \approx 0.50, 1.02, 1.13$ and 
$b_{\alpha} \approx 0.78, 0.71, 2.55$} for $\alpha=1,2,3$ respectively (see inset). Essentially, these results reveal that random fermionic Gaussian states, while lacking many of the features of generic many-body states, exhibit the same amount of nonstabilizerness resources as generic Haar states, apart from logarithmic corrections in the system size. Notably, our results, combined with those of Ref.~\cite{Turkeshi_2024_2}, suggest that under random free-fermionic time evolution, nonstabilizerness likely saturates on a timescale $\propto \log L$, approaching a value near the Haar value. In contrast, entanglement in these systems requires an exponentially longer time, scaling as $\propto L^2$~\cite{Nahum_2017}, to reach its Haar value. 
In the next Section, we provide an analytical argument to explain the observed logarithmic corrections, based on the qualitative similarity between SREs and participation entropies in the computational basis. \\
\rednewthree{Before addressing this point, we note that, as discussed above, our procedure cannot formally discriminate between pseudo-magic~\cite{Gu_2024} and genuinely magic states when restricted to a polynomial number of samples, rather than an
exponentially large one. Consequently, \emph{in principle}, the observed leading
linear scaling of the SREs for random fermionic Gaussian states could be
spurious and originate from a form of pseudo-magic, rather than reflecting a
genuinely extensive amount of magic. However, there is no clear indication that this scenario is realized. On the contrary, the analytical calculations presented in the next section for qualitatively similar participation entropies in the computational basis provide evidence for the existence of an extensive term. }

\begin{figure}[t!]
\centering
\includegraphics[width=0.9\linewidth]{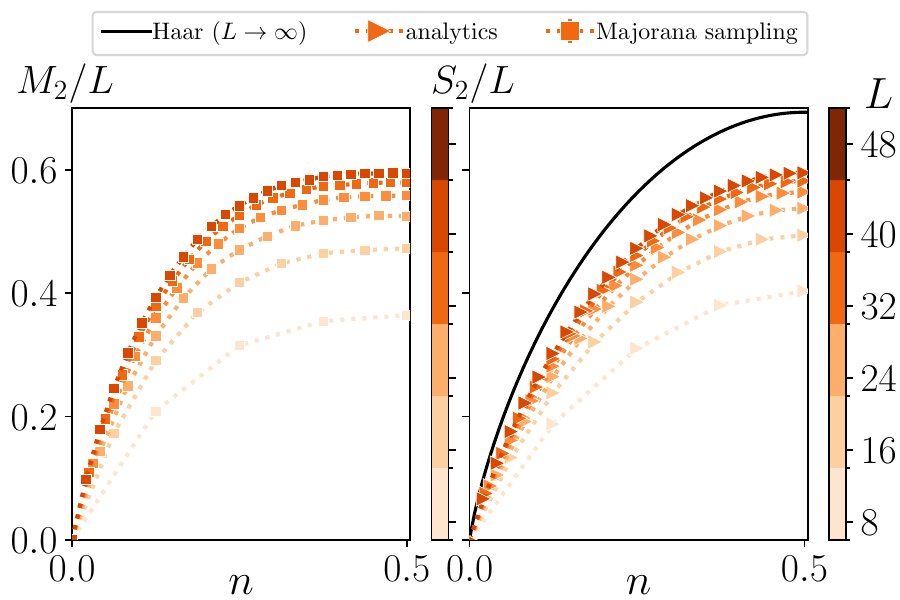}
\caption{Average SRE density $M_{2}/L$ (left) and PRE density $S_{2}/L$ (right) for random gaussian states with fixed number of particles $N$. Different system sizes $L$ are explored (see color bar), with $n$ representing the ratio $N/L$. The participation entropy $S_{2}$ is computed analytically with Eq.\eqref{eq:ipr}, while the SRE is estimated with Majorana sampling ($200$ realizations of random states in $\gausstate_{L,N}^{\text{pure}}$, with $5 \cdot 10^3$ Majorana samples for each state). The black line represents the leading extensive contribution, which is the same for both Haar-random states and fermionic Gaussian states and corresponds to the first term in Eq.\eqref{eq:participation_entropies_final}.
\label{fig:random_gaussian_2} }
\end{figure}

\paragraph{Connection with Inverse Participation 
Ratios. ---}
For simplicity, we limit the discussion to pure states, $\hat{\rho} = \ketbra{\psi}{\psi}$, and consider the overlap between $\ket{\psi}$ and a generic Fock state $\ket{\pmb{z}}$. The probability $p_{\psi}(\pmb{z}) = |\braket{\pmb{z}|\psi}|^2$, also known as formation probability~\cite{Najafi_2024, Rajabpour_2016, Rajabpour_2022, Rajabpour_2025,Rajabpour_2025_2}, quantifies the wave function's spreading over the occupation-number basis, while the characteristic probability $\pi_{\psi}(\pmb{z})$ measures its spreading over the set of Pauli operators. Therefore, strong similarities between the two quantities are expected. Analogous to the case of operators, we can define the Participation Rényi Entropies (PREs) as 
\begin{equation}
    S_{\alpha}(\rho) = \frac{1}{1-\alpha} \log I_{\alpha}(\rho) \, ,
\end{equation}
where $I_{\alpha}(\rho) = \sum_{\pmb{z}} p_{\rho}^{\alpha}(\pmb{z})$ is known as Inverse Participation Ratio (IPR).
Note that, in this case, there is no need to subtract $\log D$, as in the SRE, nor to filter out any contributions. IPRs (and PREs) are powerful tools for distinguishing phases of matter, both with and without disorder, independent of system-specific observables, and have been applied across a wide range of models and settings~\cite{Misguich_2016, Mac_2019, Evers_2008, Murphy_2011, Sierant_2022,Turkeshi_2024_1,Tirrito_2024}. 
\red{The PREs can provide bounds for the SREs~\cite{Tirrito_2025}. For example, for $\alpha=2$, it is easy to show that $M_2(\rho) \leq 2 S_2(\rho)$. Moreover, averaging the IPRs over all states in the same Clifford orbit, i.e.\ states obtained from $|\psi\rangle$ via Clifford transformations, provides direct access to the SREs~\cite{Turkeshi_2023}. For instance, the quantity $F(\rho) = I_3(\rho) - I_2^2(\rho)$, which quantifies the flatness of the distribution $p_{\psi}(\pmb{z})$, exhibits a linear scaling with $e^{-M_2(\rho)}$ when averaged over the Clifford orbit of $|\psi\rangle$~\cite{Turkeshi_2023}. In general, it is reasonable to expect qualitatively similar behavior from PREs and SREs in quantum many-body systems, as both quantify the entropy of a state's participation over a fixed basis. However, for analytical treatment, we find it more convenient to work with the PREs. In particular, below we are able to analytically demonstrate the presence of logarithmic corrections in $L$, similar to those observed for the SREs.
}\\

In case of fermionic Gaussian states with fixed number of particles $N$, the wave function amplitudes can be written as Slater determinants. Specifically, given a set of $N$ orthonormal vectors of size $L$ (orbitals) stored as columns of an $L \times N$ matrix $V$, one has 
\begin{equation}
   \braket{\pmb{z}|\psi} = \det \left[ V|_{\pmb{z}} \right] \, ,
\end{equation}
where the multi index $\pmb{z}$ select the rows of $V$ corresponding to occupied sites (i.e.\ $z_i = 1$) \footnote{Note a slight abuse of notation: here, the multi-index selects only the rows of the matrix, whereas above, a similar notation was used to select both rows and columns.}. Averaging over states $\ket{\psi} \in \gausstate_{L;N}$ is thus equivalent to averaging over random isometry matrices $V$. Specifically, one has 
\begin{equation}
   \mathbb{E}_{\psi \sim \gausstate_{L;N}}[p_{\psi}^{\alpha}(\pmb{z})] = \mathbb{E}_{U \sim \Haar}[|\det \left[ U|_{\pmb{z}}]|^{2 \alpha}  \right] \, ,
\end{equation}
where $U|_{\pmb{z}}$ is a $N \times N$ sub-matrix of an $L \times L$ unitary (Haar) matrix $U$. The statistics of eigenvalues of sub-matrices of Haar matrices is well known in random matrix theory~\cite{Zyczkowski_2000,Petz_2004,Meckes_2019}, therefore this average can be expressed explicitly as 
\begin{align}\label{eq:coulomb_gas_integral}
    \begin{split}
        &\mathbb{E}_{U \sim \Haar}[|\det \left[ U|_{\pmb{z}}]|^{2 \alpha}  \right] = \frac{1}{Z_{L,N}} \int_{|\lambda_i| < 1} d \pmb{\lambda} \, F(\pmb{\lambda}) \\
        & F(\pmb{\lambda}) = \prod_{i=1}^N |\lambda_i|^{2 \alpha}(1 - |\lambda_i|^2)^{L-N-1}  \prod_{i < j} |\lambda_i - \lambda_j|^2   \, ,
    \end{split}
\end{align}
where $\pmb{\lambda}=(\lambda_1, ... ,\lambda_N)$ are the eigenvalues of $U|_{\pmb{z}}$, the integral is over the unit disk in the complex plane and $Z_{L,N}$ is a suitable normalization constant. This expression can also be interpreted as the partition function of a gas of particles interacting through a logarithmic potential and confined by an external field~\cite{Nadal_2010,Bernard_2021,Debruyne_2023}. The integrals can be performed analytically by rewriting the full expression as a Vandermonde determinant~\cite{Debruyne_2023}. The final result is the following exact expression for the average IPRs (see Appendix~\ref{appendix2} for further details on the calculation)
\begin{align}\label{eq:ipr}
    \begin{split}
     \mathbb{E}_{\psi \sim \gausstate_{L,N}^{\text{pure}}}[I_{\alpha}(\rho)] =\binom{L}{N} \prod _{j=0}^{\alpha-1} \frac{(j+N)! (j+L-N)!}{j! (j+L)!} \, .
    \end{split}
\end{align}
The factor $\binom{L}{N}$ comes from counting the number of binary strings $\pmb{z}$ of length $L$ that contain exactly $N$ occupied sites (i.e.\ $N$ entries equal to $1$). By using the approximation $\mathbb{E}_{\psi \sim \gausstate_{L,N}^{\text{pure}}}[S_{\alpha}] \simeq \frac{1}{1-\alpha} \log \mathbb{E}_{\psi \sim \gausstate_{L,N}^{\text{pure}}}[I_{\alpha}(\rho)]$ and expanding the resulting expression for large $L$ and fixed $n = N/L$ and $\alpha$, one finds 
\begin{align}\label{eq:participation_entropies_final}
    \begin{split}
    \mathbb{E}_{\psi \sim \gausstate_{L,N}^{\text{pure}}}[S_{\alpha}] \simeq & \, L  \left(- n \log n - (1-n) \log(1-n) \right) + \\ &- \frac{1 + \alpha}{2} \log L  + c(n; \alpha) \, ,
    \end{split}
\end{align}
where $c(n; \alpha)$ is a constant which does not depend on $L$. The first term, which is extensive in $L$, represents the leading contribution and can also be derived using a saddle-point approximation of the integral Eq.~\ref{eq:coulomb_gas_integral} (similar to the approach used in  Refs.~\cite{Nadal_2010,Bernard_2021} for the entanglement entropies). The second term is instead a correction which is logarithmic in the system size $L$,
exactly as observed for the SREs (see inset in Fig.~\ref{fig:random_gaussian_1}). This correction can be interpreted as arising from fluctuations around the saddle point~\cite{Tiutiakina_2023} and has been previously observed in both the fermionic frame potential~\cite{Tiutiakina_2023} and the PREs of critical quadratic fermion systems~\cite{Rajabpour_2022}.
The similarity between PREs and SREs is evident in Fig.~\ref{fig:random_gaussian_2}, where we consider 
random number preserving fermionic Gaussian states $\ket{\psi} \in \gausstate_{L,N}^{\text{pure}}$ and we
plot the average 2-SRE, obtained numerically via Majorana sampling, alongside the analytically derived average 2-PRE.\\
Interestingly, while numerous studies have investigated IPRs in disordered systems, most have focused on the single-particle sector, with only a few addressing $I_{\alpha}$ for states of many non-interacting particles~\cite{Murphy_2011}. Notably, to the best of our knowledge, the expression Eq.~\eqref{eq:ipr} for the average IPR of free-fermionic states is derived here for the first time. \\ 

\begin{figure}[t!]
\centering
\includegraphics[width=0.9\linewidth]{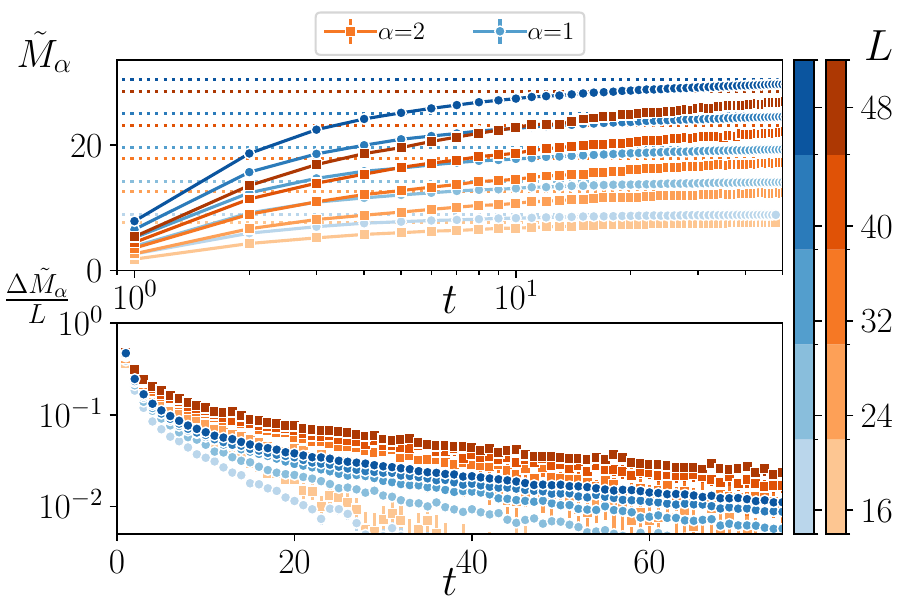}
\caption{Top: average filtered SREs $\tilde{M}_{\alpha}$ ($\alpha=1,2$) for random free-fermionic brick wall circuits (Eq.~\ref{eq:brickwall}) as a function of time $t$ \red{(particle-number is not conserved)}. Dotted horizontal lines represent saturation values $\tilde{M}_{\alpha}^{\text{sat}}$, extracted from data in Fig.~\ref{fig:random_gaussian_1}. Bottom: the difference $\Delta \tilde{M}_{\alpha}/L = (\tilde{M}_{\alpha}^{\text{sat}} - \tilde{M}_{\alpha})/L$ approaches
exponential decay. 
\label{fig:random_gaussian_circuit} }
\end{figure}

\paragraph{Random circuit dynamics. ---} After studying the magic of globally random Gaussian states, we investigate how magic evolves under local random Gaussian dynamics. Specifically, we initialize $L$ qubits in the vacuum state $\ket{0}$ and evolve them with a brick wall unitary circuit composed of random Gaussian two-qubit gates \red{that do not conserve particle number}, as follows:
\begin{equation}\label{eq:brickwall}
\includegraphics[width=0.65\linewidth, valign=c]{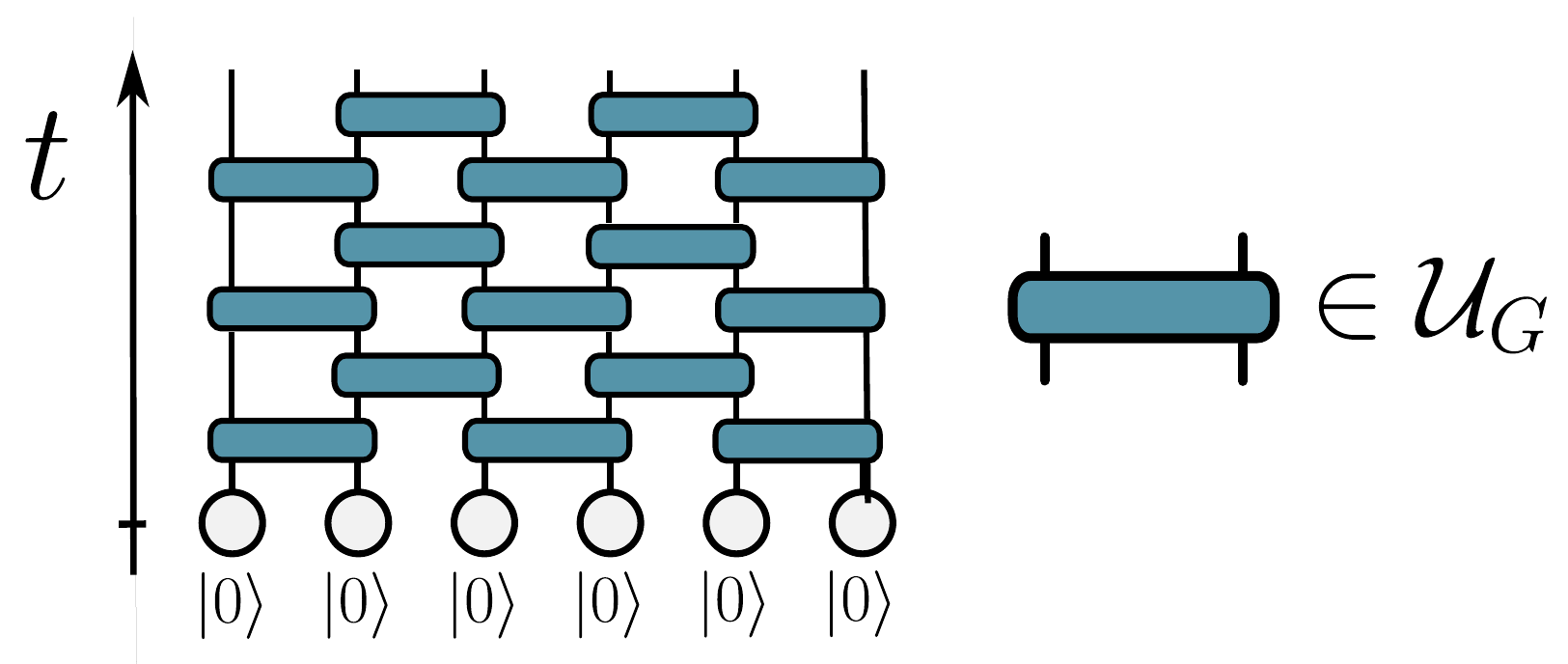}
\end{equation}
This model exhibits a `nongeneric' behavior with respect to entanglement~\cite{Nahum_2017}, as the average entanglement entropy $S_E$ behaves diffusively over time, growing as $t^{1/2}$~\cite{Rosz_2016, Nahum_2017}. This contrasts sharply with the linear growth $S_E \sim t$ observed in generic systems.
This raises the question of whether the magic of the time-evolved state in the model described by Eq.~\ref{eq:brickwall} also exhibits `nongeneric' behavior. Specifically, we can examine how magic saturates in this circuit compared to a similar circuit with local Haar gates, where saturation occurs within a time logarithmic in the system size, $t \sim \log L$, as shown in Ref.~\cite{Turkeshi_2024_2}.
To investigate this, we simulated the Gaussian dynamics for numerous random realizations of the circuit, employing Algorithm~\ref{alg:QA} to compute the magic of the time-evolved state at time $t$. We explored different system sizes $L \in [16, 48]$. The results presented in Fig.~\ref{fig:random_gaussian_circuit} show that the magic rapidly approaches its saturation value, $\tilde{M}_{\alpha}^{\text{sat}}$, which corresponds to the average value for globally random Gaussian states extracted from Fig.~\ref{fig:random_gaussian_1}. Specifically, we quantify the approach to saturation by computing the rescaled difference, $\Delta \tilde{M}_{\alpha}/L = (\tilde{M}_{\alpha}^{\text{sat}} - \tilde{M}_{\alpha})/L$. The data indicate that this quantity approaches an exponential decay, $\Delta \tilde{M}_{\alpha}/L \simeq e^{-t/\tau}$, for sufficiently large system sizes. This behavior mirrors the findings reported in Ref.~\cite{Turkeshi_2024_2} for the Haar circuit, although the characteristic time appears to be significantly larger in our case. Specifically, from our fits, we obtain $\tau \approx 45.7$ for $\alpha=1$ and $\tau \approx 43.7$ for $\alpha=2$, whereas Ref.\cite{Turkeshi_2024_2} reports a value of $\tau \simeq 2.3$ for the Haar circuit (and $\alpha=2$).
Overall, these findings suggest that a Gaussian (matchgate) circuit can efficiently generate maximally magic states in a depth $t$ that scales logarithmically with the number of qubits $L$. Notably, the fact that the unitary gates are Gaussian (matchgates) does not impose a strong constraint on the production of magic.

\begin{figure}[t!]
\centering
\includegraphics[width=0.9\linewidth]{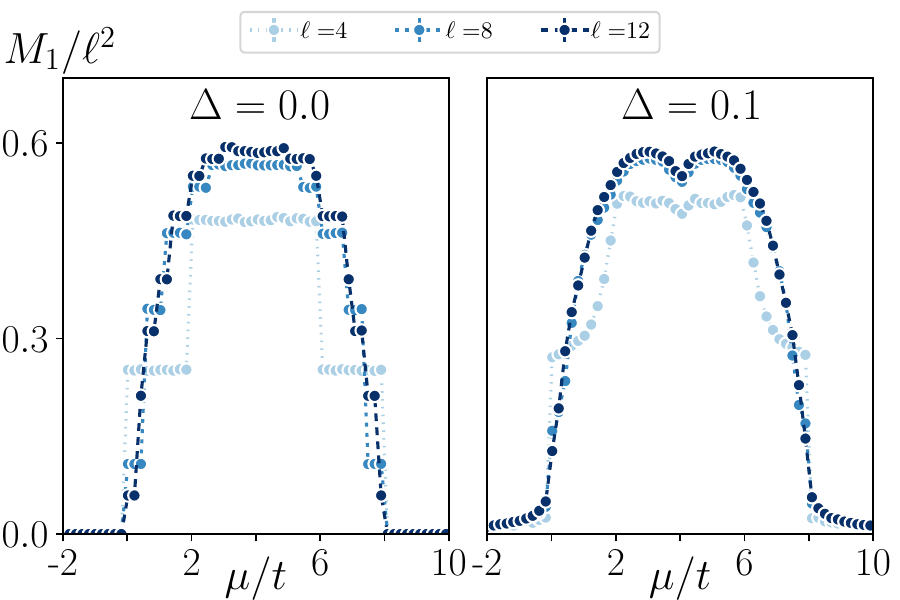}
\caption{Average SRE density $M_{1}/\ell^2$ in the ground state of the 2D Hamiltonian in Eq.(\ref{eq:2D_kitaev_H}) as a function of the chemical potential $\mu$, at $\Delta = 0$ (left) and $0.1$(right) and different lattice sizes $\ell$. The SRE is estimated using $2000$ samples per state. Error-bars are smaller than symbol size.
\label{fig:m2_2D_kitaev} }
\end{figure}

\paragraph{2D fermionic systems. ---} 
Until now, the study of magic has primarily been limited to 1D systems, as existing methods for computing the SREs (e.g.\ those proposed in Refs.~\cite{Lami_2023_2,Tarabunga_2024_2}) rely on tensor networks, which face significant limitations in higher dimensions. In fact, representing typical many-body wave functions using MPS requires scaling the bond dimension exponentially with the linear system size $\ell$. While Tree Tensor Networks sampled via Markov chain Monte Carlo~\cite{Tarabunga_2023_1} offer a possible alternative, they are hindered by the approximate nature of the state representation and autocorrelation time. The Majorana sampling method we introduced overcomes these challenges, as it is unaffected by higher dimensionality or entanglement growth. As a paradigmatic example, we examine the SREs in the ground state of a topological superconductor on a 2D square lattice of size $\ell \times \ell$. In momentum space, the Hamiltonian reads~\cite{Alicea_2012,Chamon_2014, Sato_2017}
\begin{equation}\label{eq:2D_kitaev_H}
    \hat{H} = \frac{1}{2}\sum_{\bm{k}} \hat{\Psi}_{\bm{k}}^\dagger \mathcal{H}(\bm{k}) \hat{\Psi}_{\bm{k}} \, \, , \quad \mathcal{H}(\bm{k}) = \begin{pmatrix} \epsilon_{\bm{k}} & \Delta_{\bm{k}} \\ \Delta_{\bm{k}}^* & -\epsilon_{\bm{k}} \end{pmatrix} \, ,
\end{equation}
where $\hat{\Psi}^{\dag}_{\bm{k}} = 
(\hat{c}^{\dag}_{\bm{k}}, \hat{c}_{-\bm{k}} )$, $\bm{k} = (k_x,k_y)$, $\epsilon_{\bm{k}} = -(\mu-4t) - 2t[\cos(k_x)+\cos(k_y)]$, and $\Delta_{\bm{k}} = 2i\Delta[\sin(k_x)+i\sin(k_y)]$. For $\Delta = 0$ the system presents a gapless region for \(0 \leq \mu \leq 8t\). Otherwise, it is gapped for \(\Delta \neq 0\), except for specific modes at $\mu =0,4t,8t$, where the dispersion relation develops Dirac cones at the Fermi momenta. We analyze the behavior of the SREs for the ground state of the model across various phases. Specifically, in Fig.~\ref{fig:m2_2D_kitaev}, we plot the magic density $M_1/\ell^2$ as a function of the chemical potential $\mu$ in two representative scenarios: 
\begin{itemize}
    \item $\Delta = 0$, where the transition occurs from a gapped to a gapless (non-topological) region;
    \item $\Delta = 0.1$, where the phases for $0 < \mu < 4t$ and $4t < \mu < 8t$ correspond to topological phases with opposite chiralities, separated by a double-degenerate mode at the critical point $\mu = 4t$.
\end{itemize}

For $\Delta = 0$, we observe significant finite-size effects, manifested as spurious jumps in the magic density $M_1/\ell^2$, which is expected to evolve into a smooth curve in the thermodynamic limit, with non-analytic behavior occurring only at the phase boundaries $\mu = 0, 8t$. 
These jumps result from the discreteness of the modes that contribute to the ground state as the chemical potential varies. At $\mu = 4t$, where the bands reach half-filling, the state reaches its maximal magic value. Instead, for $\mu < 0$ ($\mu > 8t$), the ground state becomes a stabilizer, as the negative (positive) single-particle energy band associated to the trivial fermions $\hat{c}_{\bm{k}}$ is fully occupied.


Turning on a finite pairing interaction $\Delta$ gives rise to interesting phenomena: 
\begin{enumerate}
    \item The magic of the ground state appears robust under the opening of a gap in the formerly critical region, still exhibiting extensive behavior. 
    \item The curve is less influenced by finite-size effects, as the ground state remains an insulator with the negative single-particle energy band associated to the Bogoliubov fermions fully occupied. In this region, no discrete jumps occur from the addition of modes to the many-body wave function.
    \item At $\mu = 4t$, the system becomes gapless, and the magic density captures this behavior through its derivative, which abruptly changes sign at the critical point. This is similar to observations in other 2D models~\cite{Tarabunga_2023_1,Falcao_2024}.
\end{enumerate}

\paragraph{Conclusions and Outlook. ---} 
We presented an efficient method for computing Stabilizer Rényi Entropies (SREs) of fermionic Gaussian states, based on a novel algorithm called Majorana sampling. This algorithm, operating within the covariance matrix formalism, enables exact sampling of Majorana monomials with probabilities proportional to the square of their expectation values. Our approach serves as a powerful tool for investigating nonstabilizerness in free-fermionic states, regardless of their entanglement (which in general exhibits a volume-law scaling).
Through a detailed numerical analysis of random fermionic Gaussian states, we have demonstrated that these states exhibit nonstabilizerness features akin to those of generic Haar-random states, with only subleading corrections that scale logarithmically with the system size. 
We substantiate these findings with analytical calculations for related quantities—specifically, the Inverse Participation Ratio (IPR) and the Participation Rényi Entropies (PREs) in the computational (or Fock) basis. We derive an exact formula for the average values of these quantities over random fermionic Gaussian states, showing similar logarithmic corrections.
Overall, our results demonstrate that matchgate circuits, despite carrying no resource if considered within the resource theory of Gaussian states, are highly effective at generating and propagating nonstabilizerness. 
As an intriguing physical scenario, we applied our method to evaluate the magic in the ground-state of a two-dimensional non-interacting system of fermions. Here, topological properties are observed to influence the nonstabilizerness density of the system.\\
Looking ahead, several exciting avenues remain to be explored. One promising direction is to extend our approach to more complex systems in higher dimensions, where the interplay between entanglement and nonstabilizerness could uncover new insights into the phases of matter. Our method could also provide insights into the non-equilibrium dynamics of quantum magic in novel setups, where nonstabilizerness and quantum measurements interplay to give rise to new out-of-equilibrium phases of matter~\cite{Paviglianiti_2024, Fux_2024, Bejan_2024_1}. 
Additionally, the Majorana sampling approach could be used to explore the distribution associated to a state over the operators in various ways, such as by analyzing the distribution of operator lengths.
In the end, our results suggest a promising way to understand the complex structure of quantum states and their complexity in many different physical systems, even when these states are just simple Gaussian fermionic states.\\

\paragraph{Acknowledgments. ---} 

We are especially grateful to Xhek Turkeshi for his valuable suggestions on the filtered stabilizer Rényi entropies. We also thank Andrea De Luca for his valuable suggestions on the IPR calculation, and Emanuele Tirrito, Lorenzo Leone, Piotr Sierant and Marcello Dalmonte for their inspiring discussions. This work was supported by ANR-22-CPJ1-0021-01, the ERC Starting Grant 101042293 (HEPIQ), the PNRR MUR project PE0000023-NQSTI, and the PRIN 2022 (2022R35ZBF) - PE2 - ``ManyQLowD''.  \\

\bibliography{bib}
\bibliographystyle{quantum}
\clearpage
\setcounter{section}{0}
\setcounter{secnumdepth}{2}

\onecolumngrid
\begin{center}
    \textbf{End Matter}
\end{center}

\section{Details on the sample based estimation}\label{appendix1}
In this section, we outline the process of estimating the Stabilizer Rényi Entropies (SREs) and filtered Stabilizer Rényi Entropies (filtered SREs) for a state $\hat{\rho}$ after obtaining a batch of samples $\mathcal{X}=\{ \pmb{x}^{(i)} \}_{i=1}^{\nsamp}$ via Majorana sampling (Algorithm~\ref{alg:QA}). By construction, samples are distributed according as $\pmb{x}^{(i)} \sim \pi_{\rho}$. The algorithm also outputs the probability associated of each sample $\{ \pi_{\rho}(\pmb{x}^{(i)}) \}_{i=1}^{\nsamp}$. To instead sample from $\tilde{\pi}_{\rho}$, it is sufficient to exclude from $\mathcal{X}$ all occurrences of the identity $\hat{I}=\Id_1 \dots \Id_L$ and of the parity $\parity = \PauliZ_1 \dots \PauliZ_L$. Note that these Pauli strings correspond to the bit configurations $\pmb{x}_0 = (0,0,\dots,0)$ and $\pmb{x}_1 = (1,1,\dots,1)$, respectively. Note also that for large system sizes, $L \gg 1$, it is generally very unlikely to sample one of these strings, as both have exponentially small probabilities: $\pi_{\rho}(I)=\pi_{\rho}(\parity)=D^{-1}$. In any case, after filtering away $I$ and $\parity$, one obtains a restricted batch of samples $\tilde{\mathcal{X}} = \{ \pmb{x}^{(i)} \}_{i=1}^{\tilde{\nsamp}}$, with $\tilde{\nsamp} \leq \nsamp$ and $\pmb{x}^{(i)} \neq \pmb{x}_0, \pmb{x}_1$. 
Now we set
\begin{equation}\label{eq:def_Q}
        M_{\alpha}(\rho) =  \frac{1}{1-\alpha} \log ( Q_{\alpha} ) - \log D \, ,\qquad 
    \tilde{M}_{\alpha}(\rho) =  \frac{1}{1-\alpha} \log ( \tilde{Q}_{\alpha}) - \log D \, ,
\end{equation}
which defines the quantities $Q_{\alpha}, \tilde{Q}_{\alpha}$, to be estimated from the samples as $Q_{\alpha}^{\text{est}}, \tilde{Q}_{\alpha}^{\text{est}}$. For the filtered SREs, we have
\begin{equation}
    \tilde{M}_{\alpha}(\rho) = \frac{1}{1-\alpha} \log \left( \sum_{\pmb{x}} \tilde{\pi}_{\rho}^{\alpha}(\pmb{x}) \right) - \log (D -2) =  \frac{1}{1-\alpha} \log \left( \sum_{\pmb{x}} \tilde{\pi}_{\rho}(\pmb{x}) \pi_{\rho}^{\alpha-1}(\pmb{x}) \right) - \log D \, , 
\end{equation}
where we used $\tilde{\pi}_{\rho}(\pmb{x}) = \pi_{\rho}(\pmb{x}) \frac{D}{D-2}$, $\forall \pmb{x} \neq \pmb{x}_0, \pmb{x}_1$.
Therefore $\tilde{Q}_{\alpha} = \sum_{\pmb{x}} \tilde{\pi}_{\rho}^{\alpha}(\pmb{x}) \pi_{\rho}^{\alpha-1}(\pmb{x})$, and the corresponding estimator is given by
\begin{equation}
    \tilde{Q}_{\alpha}^{\text{est}} = \frac{1}{\tilde{\nsamp}} \sum_{i=1}^{\tilde{\nsamp}} \pi_{\rho}^{\alpha-1}(\pmb{x}^{(i)}) \, .
\end{equation}
For the original SREs instead, we have (restoring the contributions of $I$ and $\parity$)
\begin{equation}
    M_{\alpha}(\rho) = \frac{1}{1-\alpha} \log \left( \sum_{\pmb{x}} \pi_{\rho}^{\alpha}(\pmb{x}) \right) - \log D =  \frac{1}{1-\alpha} \log \left( \frac{2}{D^{\alpha}} + \frac{D-2}{D} \sum_{\pmb{x}} \tilde{\pi}_{\rho}(\pmb{x}) \pi_{\rho}^{\alpha-1}(\pmb{x}) \right) - \log D \, ,
\end{equation}
which implies $Q_{\alpha} = \frac{2}{D^{\alpha}} + \frac{D-2}{D} \sum_{\pmb{x}} \tilde{\pi}_{\rho}(\pmb{x}) \pi_{\rho}^{\alpha-1}(\pmb{x})$. In this case the estimator is therefore
\begin{equation}
    Q_{\alpha}^{\text{est}} = \frac{2}{D^{\alpha}} + \frac{D-2}{D} \frac{1}{\tilde{\nsamp}} \sum_{i=1}^{\tilde{\nsamp}} \pi_{\rho}^{\alpha-1}(\pmb{x}^{(i)}) \, .
\end{equation}
For the particular case $\alpha=1$, in which the Rényi entropies reduce to the Shannon entropy, we have instead 
$\tilde{M}_{1}(\rho) = Q_{1} - \log D$, $M_{1}(\rho) = \tilde{Q}_{1} - \log D$, with 
\begin{equation}
    \tilde{Q}_{1} =  - \sum_{\pmb{x}} \tilde{\pi}_{\rho}(\pmb{x}) \log \pi_{\rho}(\pmb{x}) \, , \qquad  Q_{1} =  - \frac{2}{D} \log \frac{1}{D} - \frac{D-2}{D} \sum_{\pmb{x}} \tilde{\pi}_{\rho}(\pmb{x}) \log \pi_{\rho}(\pmb{x}) \, .
\end{equation}
Therefore the estimation goes as follows
\begin{equation}
    \tilde{Q}_{1}^{\text{est}} = \frac{1}{\tilde{\nsamp}} \sum_{i=1}^{\tilde{\nsamp}} \left( - \log \pi_{\rho}(\pmb{x}^{(i)}) \right)
\qquad 
    Q_{1}^{\text{est}} =  \frac{D-2}{D} \frac{1}{\tilde{\nsamp}} \sum_{i=1}^{\tilde{\nsamp}} \left( - \log \pi_{\rho}(\pmb{x}^{(i)}) \right) + \frac{2}{D} \log D \, .
\end{equation}

\rednewnew{Notice that while $\tilde{Q}_{\alpha}^{\text{est}}$, $Q_{\alpha}^{\text{est}}$ are always unbiased estimators for, respectively, $\tilde{Q}_{\alpha}$ and $Q_{\alpha}$, when $\alpha>1$ the logarithm one has to take in Eq.\eqref{eq:def_Q} to estimate possibly induces a bias in the estimation. Specifically, 
\begin{equation}
        M_{\alpha}^{\text{est}}(\rho) =  \frac{1}{1-\alpha} \log ( Q_{\alpha}^{\text{est}} ) - \log D \, ,\qquad 
    \tilde{M}_{\alpha}^{\text{est}}(\rho) =  \frac{1}{1-\alpha} \log ( \tilde{Q}_{\alpha}^{\text{est}}) - \log D \, ,
\end{equation}
are not unbiased estimators for $M_{\alpha}, \tilde{M}_{\alpha}$
(this problem is not present for the case $\alpha=1$). One can correct the bias by Taylor expanding the logarithm, and obtaining the following correction
\begin{align}\label{eq:corr_est}
    \begin{split}
     M_{\alpha}^{\text{corr.est}}(\rho) &=  \frac{1}{1-\alpha} \left( \log ( Q_{\alpha}^{\text{est}} )  - \frac{1}{2} \text{BIAS}_{\alpha} \right)- \log D  \\ 
    \tilde{M}_{\alpha}^{\text{corr.est}}(\rho) &=  \frac{1}{1-\alpha} \left( \log ( Q_{\alpha}^{\text{est}} )  - \frac{1}{2} \tilde{\text{BIAS}}_{\alpha} \right) - \log D \, ,
    \end{split}
\end{align}
where 
\begin{equation}
\text{BIAS}_{\alpha} = \frac{\Var[Q_{\alpha}^{\text{est}}]} {(Q_{\alpha}^{\text{est}})^2} \qquad \tilde{\text{BIAS}}_{\alpha} = \frac{\Var[\tilde{Q}_{\alpha}^{\text{est}}]}{(\tilde{Q}_{\alpha}^{\text{est}})^2} \, .
\end{equation}
While the variance of the estimators is not known a priori, it can nevertheless be estimated from the data via the sample variance. To obtain the data presented in 
Fig.\ref{fig:random_gaussian_1} we indeed used the corrected estimators given in Eq.\ref{eq:corr_est}. In Fig.\ref{fig:bias} (upper panel), we plot the values of $\tilde{\text{BIAS}}_{\alpha}$ for $\alpha=2,3$ as a function of the system size $L$. While for $\alpha=2$ the bias remains always modest, i.e. below $\sim 0.07$, for $\alpha=3$ it grows much more, up to almost $0.5$ for large $L$. This behavior reflects the difficulty in estimating well the exponentially small value of $\tilde{Q}_{\alpha}$ with a limited number $\nsamp$ of samples.
While these values could suggest that our estimation of the nonstabilizerness is not precise, this is not the case. In fact, normalizing the bias by the leading contribution, i.e.\ computing $\tilde{\text{BIAS}}_{\alpha} / \log \left( \tilde{Q}_{\alpha}^{\text{est}} \right)$,
we find that this quantity remains small ($< 0.01$), even for $\alpha = 3$ and large $L$ (see lower panel in Fig.~\ref{fig:bias}). This shows that our estimation is indeed accurate. In a nutshell, although estimating the exponentially small $\tilde{Q}_{\alpha}$ with a small relative error can be hard for $\alpha > 1$ and large $L$, this can have little impact on the \emph{logarithm} of these quantities. Indeed, because $\tilde{Q}_{\alpha}$ is already very small, even large relative errors in $\tilde{Q}_{\alpha}$ result in only minor relative deviations in $\log \tilde{Q}_{\alpha}$.
\begin{figure}[t!]
\centering
\includegraphics[width=0.45\linewidth]{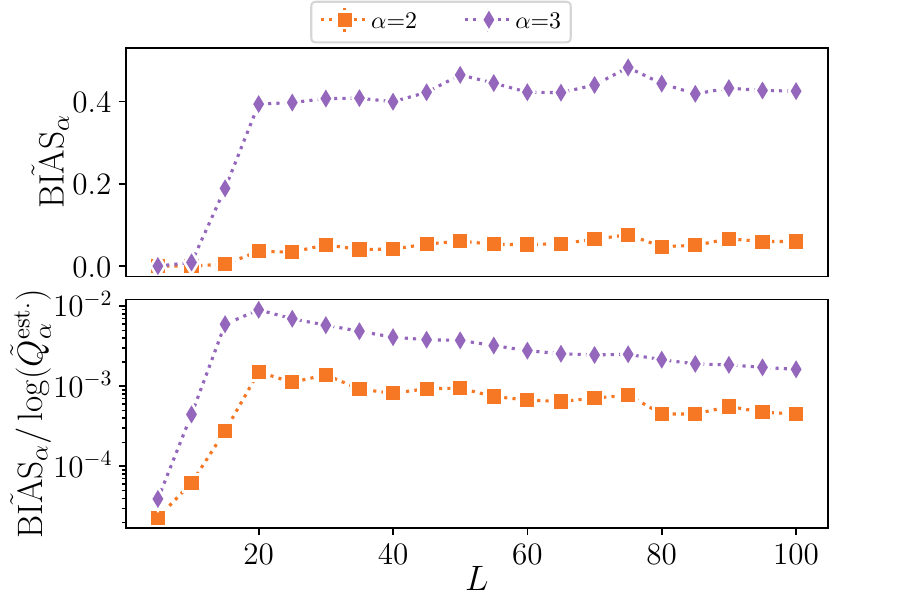}
\caption{Estimator bias for Renyi index $\alpha=2,3$ and different values of the system size $L$. Data refer to Fig.\ref{fig:random_gaussian_1}. In the lower panel the bias is normalized with the leading term $\log \left( \tilde{Q}_{\alpha}^{\text{est}} \right)$, showing that the relative correction is small.  
\label{fig:bias}
 }
\end{figure}
}

\section{Further details on the Inverse Participation Ratio calculation}\label{appendix2}
Computing the average Inverse Participation Ration (IPR) for a random fermionic Gaussian state with a fixed particle number $N$ reduces to evaluating $\mathbb{E}_{U \sim \Haar}[|\det(U|_{\pmb{z}})|^{2\alpha}]$, where $U|_{\pmb{z}}$ is an $N \times N$ submatrix of a unitary Haar matrix $U$ of size $L \times L$. Leveraging the known measure of the eigenvalues of a Haar submatrix, one obtains Eq.\eqref{eq:coulomb_gas_integral}. Now, the idea is to rewrite the integrand 
\begin{equation}
\prod_{i < j} |\lambda_i - \lambda_j|^2 \prod_{i=1}^N |\lambda_i|^{2 \alpha}(1 - |\lambda_i|^2)^{L-N-1} 
\end{equation}
as a determinant of a suitable matrix. To this purpose we first use the Vandermonde identity to obtain
\begin{equation}
    \prod_{i < j} |\lambda_i - \lambda_j|^2 = \prod_{i < j} (\lambda_i - \lambda_j) \prod_{i < j} (\lambda_i^* - \lambda_j^*) = \det_{1 \leq i,j \leq N} [\lambda_i^{j-1}] \det_{1 \leq i,j \leq N} [(\lambda_i^*)^{j-1}] \, .
\end{equation}
Now, we exploit the following version of the Cauchy-Binet identity~\cite{Debruyne_2023} 
\begin{equation}
  \int dx_1 ... dx_N \det_{1 \leq i,j \leq N}[f_i(x_j)] \det_{1 \leq i,j \leq N}[g_i(x_j)] \prod_{i=1}^N h(x_i) = N! \det_{1 \leq i,j \leq N}\big[\int dx f_i(x) g_j(x) h(x) \big] \, ,
\end{equation}
with $f_i(\lambda) = \lambda^{i-1}$, $g_j(\lambda) = (\lambda^*)^{j-1}$ and $h(\lambda) = |\lambda|^{2 \alpha}(1 - |\lambda|^2)^{L-N-1}$. We thus obtain 
\begin{equation}\label{eq:final_det}
  \int_{|\lambda_i| < 1} d \lambda_1 ... d \lambda_N \prod_{i < j} |\lambda_i - \lambda_j|^2 \prod_{i=1}^N |\lambda_i|^{2 \alpha}(1 - |\lambda_i|^2)^{L-N-1}  = N! \det_{1 \leq i,j \leq N}\big[ \int_{|\lambda| < 1} d\lambda \, \lambda^{i-1} (\lambda^*)^{j-1} |\lambda|^{2 \alpha}(1 - |\lambda|^2)^{L-N-1} \big] \, .   
\end{equation}
Switching to polar coordinates $\lambda = r e^{i \theta}$, with $r \in [0,1]$, the integral becomes
\begin{align}
    \begin{split}
         \int_{0}^{1} dr \int_{0}^{2 \pi} d\theta  \, r^{i+j-1} e^{i \theta (i-j)} r^{2\alpha}(1 - r^2)^{L-N-1} = 2 \pi \delta_{i,j}  \int_{0}^{1} dr \,   r^{2j-1} r^{2\alpha}(1 - r^2)^{L-N-1} \, .
    \end{split}
\end{align}
At this point, the determinant in Eq.\eqref{eq:final_det} simplifies to a straightforward computation, as the matrix is diagonal. Moreover, the normalization constant $Z_{L,N}$ is restored by evaluating Eq.\eqref{eq:final_det} at $\alpha = 0$. Combining these results, we obtain
\begin{equation}
         \mathbb{E}_{U \sim \Haar}[|\det \left[ U|_{\pmb{z}}]|^{2 \alpha}  \right] =\prod_{j=1}^N  \frac{\int_{0}^{1} dr \, r^{2j - 1 + 2\alpha}(1 - r^2)^{L-N-1}}{\int_{0}^{1} dr \, r^{2j - 1}(1 - r^2)^{L-N-1}} \, .
\end{equation}
All the integrals can be evaluated analytically, resulting in 
\begin{equation}
    \mathbb{E}_{U \sim \Haar}[|\det \left[ U|_{\pmb{z}}]|^{2 \alpha}  \right] =\prod_{j=1}^N \frac{\Gamma (j+\alpha) \Gamma (j+L-N)}{\Gamma (j) \Gamma (j+L-N+\alpha)} \, .
\end{equation}
Expanding the product and applying successive telescopic simplifications, we arrive at the final expression
\begin{equation}
    \mathbb{E}_{U \sim \Haar}[|\det \left[ U|_{\pmb{z}}]|^{2 \alpha}  \right] = \prod _{j=1}^{\alpha } \frac{\Gamma (j+N) \Gamma (j+L-N)}{\Gamma (j) \Gamma (j+L)} \, .
\end{equation}
Finally, we obtain Eq.~\ref{eq:ipr} for the IPR $I_{\alpha}(\rho) = \sum_{\pmb{z}} p_{\rho}^{\alpha}(\pmb{z})$ by introducing the factor $\binom{L}{N}$, which counts the number of binary strings $\pmb{z}$ of length $L$ with $N$ occupied sites. By expanding $\mathbb{E}_{\psi \sim \gausstate_{L,N}^{\text{pure}}}[S_{\alpha}] \simeq \frac{1}{1-\alpha} \log \mathbb{E}_{\psi \sim \gausstate_{L,N}^{\text{pure}}}[I_{\alpha}(\rho)]$ for large $L$ and fixed $n = N/L$ and $\alpha$, we find 
Eq.~\ref{eq:participation_entropies_final}. The constant $c(n;\alpha)$ is given by 
\begin{equation}
    c(n;\alpha)= \frac{1}{2 (\alpha -1)} \left( -\alpha \log (2 \pi )+2 \log (G(\alpha +1))-\alpha ^2 \log (n (1-n))\right)
\end{equation}
with $G$ the Barnes G-function.

\end{document}